\documentstyle[12pt,epsfig,rotating]{article}
\begin{document}
\hfill
\parbox{5.5cm}{{\large HU-SEFT R 1996-19} \\
              {\large 27 September 1996}} \\

\vskip 1.0cm
%

\large
\centerline {\bf B-physics potential of ATLAS:
an update}
\normalsize
 
\vskip 2.0cm
\centerline {P. Eerola~\footnote{On leave of absence from SEFT and University
of Helsinki, Helsinki, Finland}}
\centerline {\it CERN, Geneva, Switzerland}
\vskip 0.5cm
\centerline {For the ATLAS Collaboration}
\vskip 4.0cm
 
\centerline {\bf Abstract}
\vskip 1.0cm
The B-physics potential of the ATLAS experiment at LHC is described.
Simulation results are shown for the measurement of $\sin 2 \beta$,
with an emphasis on new tagging techniques. 
Other CP-violation measurements are described briefly. 
New limits are shown for the reach of
the $x_{\mathrm s}$-measurement, resulting
from increased statistics and improved fitting methods.
Some rare decay modes of B-mesons can be easily seen in ATLAS.
Analyses of channels $\mathrm{B\rightarrow \mu^+\mu^- (X)}$
are presented here.
\vskip 2.0cm
\centerline {\it Presented in BEAUTY'96, June 17-21 1996, Rome, Italy.}
\centerline {\it To be published in Nucl. Instrum. and Methods A.}
\vfill

\newpage
\section{B physics in ATLAS}
LHC is a B-factory. At $\sqrt{s}$ = 14 TeV, the cross-section for 
${\rm b\bar{b}}$ production is about 500~$\mu$b, with
$\sigma({\rm b\bar{b}})/\sigma({\rm tot}) \simeq $0.7\%.
At the start-up of LHC, the luminosity is expected to be
${\cal L} = 10^{33}$ cm$^{-2}$s$^{-1}$, corresponding to an integrated 
luminosity of
$10^4$ pb$^{-1}$ per year giving $5 \cdot 10^{12} {\rm b\bar{b}}$
pairs.

ATLAS can measure precisely CP-violation in the decay
${\rm B^0_d \rightarrow J/\psi K^0_S}$, which measures the $\beta$-angle
in the unitarity triangle, and
${\rm B^0_s}$-mixing, which measures the length of the side opposite to 
$\gamma$-angle. With these two measurements,
the triangle is already completely defined according to the Standard Model.
In order to test the Standard Model, one needs additional experimental
(over)constraints. ATLAS can contribute to measurements of the other two 
angles, $\alpha$ and $\gamma$. In addition,
rare decays provide us with an independent Standard Model test.

The ATLAS B-physics reach has been previously studied in~\cite{tp}
and in~\cite{beauty}. These publications also include detailed
descriptions of the ATLAS detector and formulae for CP-violation
measurements. Here, we concentrate on new simulation
results. New tagging techniques for the measurement of $\sin 2 \beta$
are described in detail. 
An improved limit for the $x_{\rm s}$-reach is obtained, due to 
increased signal statistics (decay channels 
$\mathrm{ B_s^0 \rightarrow  D_s^- \pi^+}$ and
$\mathrm{ B_s^0 \rightarrow  D_s^- a_1^+}$),
and refined fitting methods. 
New analyses on rare decays $\rm{B \rightarrow
\mu^+\mu^-(X)}$ are presented in some detail in this paper.

\section{The ATLAS experiment}

ATLAS is a general purpose experiment at LHC~\cite{tp}. The
new particle searches at the nominal LHC luminosity 
(${\cal L} = 10^{34}$ cm$^{-2}$s$^{-1}$) have defined most 
of the performance specifications of the detector.
B-physics requirements have been accommodated, however, in particular
in the design of the inner detector and the trigger systems.

The inner detector, located in a 2 T solenoidal field,
consists of a semiconductor tracker (SCT), providing at least
six high-precision space points over pseudorapidity range of $\pm$~2.5, and
a transition radiation tracker (TRT) providing on the average 
35 two-dimensional measurement points over $|\eta|<2.5$. Electrons 
with $p_{\mathrm T} > 1$~GeV/$c$ can be identified by the 
transition radiation they produce in the TRT. 
For the initial low-luminosity operation of LHC,
ATLAS will also be equipped with a special vertexing layer next to the
beam pipe, either a pixel or a double-sided strip layer. 
The nominal resolutions of the strip layer are
$\sigma_{\rm{IP}} = (13 \oplus 58/(p_{\mathrm T}\sqrt{|\sin \theta|}))$~$\mu$m, 
$\sigma_{\rm{z}} =(37\oplus 62/(p_{\mathrm T}\sqrt{|(\sin \theta)^3|}))$~$\mu$m,
where $\sigma_{\rm{IP}}$ is the resolution in the transverse plane, 
$\sigma_{\rm{z}}$ is the resolution in z,
$p_{\mathrm T}$ is given in GeV/$c$, and $\theta$ is the polar angle.

Muons with $p_{\mathrm T} > 5$~GeV/$c$ are identified in the muon spectrometer, 
which is a large toroid-system equipped with
muon chambers and dedicated trigger chambers. In addition, the last compartment
of the barrel hadron calorimeter can be used for identifying muons with
$p_{\mathrm T} > 3$~GeV/$c$. The low-$p_{\mathrm T}$ electron identification
used here
relies on the TRT, but preliminary studies show that electrons in b-jets
can also be identified in the Liquid-argon calorimeter with $p_{\mathrm T}$
values as low as 2~GeV/$c$~\cite{fabiola}.

The B-physics Level-1 trigger is a single muon trigger with 
$p_{\mathrm T} > 6$~GeV/$c$, $|\eta|<2.2$. At Level-1, 
the total rate is about 8 kHz, 
with 60\% b-purity (60\%~b$\rm{\bar{b}}$,
20\%~c$\rm{\bar{c}}$, 20\%~$\pi$/K decays). 
The $p_{\mathrm T}$ measurement of the trigger muon will be refined in 
Level-2. In addition, more elaborate signatures will
be used at Levels 2 and 3, requiring for example additional leptons, 
additional high-$p_{\mathrm T}$ tracks, or making mass cuts.

\boldmath
\section{Measurement of $\sin 2 \beta$}
\unboldmath

The CP-violation parameter $\sin 2 \beta$ is measured from the asymmetry
of the decays \\
${\rm{ (B^0_d \rightarrow J/\psi K^0_S)}}$ and
${\rm{ ( \bar{B}^0_d \rightarrow J/\psi K^0_S)}}$.
To distinguish B- and $\bar{\rm B}$-decays, a tag is used. 
In ATLAS, the tag can be
charge of the lepton, which originates from a semileptonic 
decay of the other B, b$\rightarrow \ell^-$. We have also investigated
B$^{**}$-tagging, in which the charge of the pion next to the signal-B
provides the tag, and
jet charge tagging, in which the charge of the jet containing the signal-B, 
or the charge of the other B-jet, or a combination of the two, provides the tag.

The asymmetry is reduced by wrong tags (dilution factor $D_{\mathrm{tag}} 
= 1-2W_{\mathrm{tag}}$, $W_{\mathrm{tag}}$~=~wrong tag fraction)
and background (dilution factor 
$D_{\mathrm{back}}= N_{\rm S}/(N_{\rm S}+N_{\rm B})$).
If a time-independent measurement of the asymmetry is performed, the 
time-integration produces an effective dilution factor 
$D_{\mathrm{int}}$ = $1/(1+x_{\rm d}^2)[\sin \Delta m t_0 +
x_{\rm d} \cos\Delta m t_0]$, where $x_{\rm d}=\Delta m/\Gamma$ 
is the mixing parameter, and $t_0$ is the starting point for 
the time-integration. With a time-dependent measurement of the asymmetry,
a better precision can be obtained, with a dilution factor
$D_{\mathrm{time}}$ from the fitting procedure.
In addition, in proton-proton collision, a small production asymmetry 
$A_{\mathrm{p}}$ of
B and $\bar{\rm B}$ mesons (${\cal O}(1\%)$) is expected.

\subsection{Lepton tagged sample}

We have three lepton tagged signal samples: 
1) muon tagged $\mathrm{J/\psi K^0_S}$, J/$\psi \rightarrow \mu^+\mu^-$,\\
2) muon tagged $\mathrm{J/\psi K^0_S}$, J/$\psi \rightarrow \mathrm{e^+ e^- }$,
3) electron tagged $\mathrm{J/\psi K^0_S}$, J/$\psi \rightarrow \mu^+\mu^-$.
The tag-electron was required to have $p_{\mathrm T} > 5$~GeV/$c$.
One of the muons must satisfy the single muon trigger.

The analysis is the same as in~\cite{tp}. The data samples were simulated
with PYTHIA, using the old default parton distribution function EHLQ1.
The ${\rm {K^0_S}}$ reconstruction efficiency has been studied in ATLAS
using a full GEANT simulation~\cite{k0s}. 
The reconstruction efficiency was found to be
95\% with a fake ${\rm {K^0_S}}$ rate of 6\%.
The J/$\psi$ reconstruction has been documented in~\cite{tp,igor}.
In addition to  $p_{\mathrm T}$ cuts and lepton identification,
secondary vertex and mass cuts are used~\cite{tp,beta}.
The reconstructed B-lifetime is required to be greater than~0.5~ps and
the $p_{\rm T}$ of the B greater than 5 GeV/$c$.

For the decay $\rm{ B \rightarrow J/\psi K^0_S \rightarrow \mu^+\mu^- K^0_S}$,
the dominant background consists of a real J/$\psi$ from a 
B decay, combined with a ${\rm {K^0_S}}$ from the fragmentation, assuming 
a hadron rejection factor of 50 in the muon spectrometer.
For the decay $\rm{ B \rightarrow J/\psi K^0_S \rightarrow e^+e^- K^0_S}$, 
the dominant background comes from an (eh)-pair
faking a J/$\psi$, combined with any
${\rm {K^0_S}}$, assuming a hadron rejection factor of 30 in the TRT. 
The fake J/$\psi$ background in the electron channel is
due to the low $p_{\rm T}$ cut of 1~GeV/$c$ for electrons, and the
large J/$\psi$ mass window (600~MeV/$c^2$) due to bremsstrahlung.
The fake J/$\psi$ background from (eh)-pairs
was not considered in~\cite{tp}. The final
results do not change significantly, however.

The final number of signal events is 21,120 for an integrated luminosity
of $10^4$~pb$^{-1}$. 
The breakdown of the signal
and background events in the different event classes is given in 
Table~\ref{table1}. The result with the lepton tagged sample is
$\delta (\sin 2 \beta) (\mathrm{stat.})$ = 0.021 (time-int.).

\begin{table}[htb]
\centering
\begin{tabular}{|l|c|c|}
\hline
Event class & $\mu$-tags & e-tags\\ \hline
$N_{\rm S}({\rm J/\psi \rightarrow e^+e^- \oplus K^0_S})$  & 9,540 & \\
$N_{\rm S}({\rm J/\psi \rightarrow \mu^+\mu^- \oplus K^0_S})$ & 6,220 & 
                                 5,360 \\ \hline
$N_{\rm B}({\rm {real \, J/\psi \rightarrow e^+e^- \oplus \pi^+\pi^-}})$ & 
               1,170 & \\
$N_{\rm B}({\rm {fake \, e^+e^- \oplus \pi^+\pi^-}})$ & 
               1,110 & \\
$N_{\rm B}({\rm {fake \, eh \oplus \pi^+\pi^-}})$ & 
               1,550 & \\ \hline
$N_{\rm B}({\rm {real \, J/\psi \rightarrow \mu^+\mu^- \oplus \pi^+\pi^-}})$ & 
               80 & 160 \\ 
$N_{\rm B}({\rm {fake \,  \mu^+\mu^- \oplus \pi^+\pi^-}})$ & 
               40 & 40 \\ 
$N_{\rm B}({\rm {fake \, \mu h \oplus \pi^+\pi^-}})$ & 
               4 & 2 \\ \hline
\end{tabular}
\caption{Number of signal and background events in the lepton tagged
${\rm{B^0_d \rightarrow J/\psi K^0_s}}$ samples.}
\label{table1}
\end{table}

Systematic uncertainties can be 
controlled using data from 
${\rm{B^+ \rightarrow J/\psi K^+}}$ and \\
${\rm{B^0_d \rightarrow J/\psi K^{0*}}}$~\cite{syst}:
with an integrated luminosity of 
$10^4$~pb$^{-1}$, 
$\delta D_{\mathrm{tag}}/ D_{\mathrm{tag}} = 0.005 $ (lepton tags),
$\delta A_{\mathrm{p}} = 0.007 $.
With decreased statistical errors in the control samples, these 
systematic uncertainties will decrease.
For ${\rm{B^0_d \rightarrow J/\psi K^0_S}}$,
$ \delta D_{\mathrm{back}}/ D_{\mathrm{back}} = 0.011 $ (estimate).
Combining these sources of systematic uncertainties, the following 
estimate is obtained:
$\delta (\sin 2 \beta)$ (syst.)= 
$\pm 0.012 \cdot \sin 2 \beta \oplus 0.011$.

\subsection{B$^{**}$ and jet charge tagging}
B$^{**}$ is a common label for orbitally excited $L=1$ state B's. 
Theoretically, one expects a narrow and a wide doublet~\cite{eichten}. 
B$^{**}$ states have been observed at LEP~\cite{b**}.
The expected properties of charged B$^{**}$ states are tabulated 
in Table~\ref{table2}.

\begin{table}[htb]
\centering
\begin{tabular}{|l|c|c|c|c|l|}\hline
Label & $j_q=L+S_q$ & $J^P$ & Mass [GeV/$c^2$] & Width [MeV/$c^2$] & 
                               Decays to \\ \hline
$\mathrm{B}_1^+$ & $3/2$ & $1^+$ & 5.742 & 21 & B$^{*0}\pi^+$ \\
$\mathrm{B}^{*+}_2$ & $3/2$ & $2^+$ & 5.754 & 25 & B$^0\pi^+$, B$^{*0}\pi^+$ 
                                                        \\ \hline
$\mathrm{B}^{*+}_0$ & $1/2$ & $0^+$ & 5.630 & 100 & B$^0\pi^+$ \\
$\mathrm{B}^{*+}_1$ & $1/2$ & $1^+$ & 5.642 & 100 & B$^{*0}\pi^+$ \\ \hline
\end{tabular}
\caption{The charged B$^{**}$ states.
Masses and widths come from theoretical estimates and
ALEPH data (unpublished).}
\label{table2}
\end{table}

The charge of the pion is a signature of the charge of the b quark:
${\rm{B^{**+} \rightarrow B^{0(*)} \pi^+ }}$,
${\rm{B^{**-} \rightarrow \bar{B}^{0(*)} \pi^- }}$,
where B$^{0(*)}$ denotes either a B$^0$ or a B$^{*0}$.
Even without a resonant structure, there is a correlation between the 
charge of pions nearby the B in phase space, and the flavour of the B.
This correlation can be used to tag B's on a statistical basis by using 
the jet charge as a tag. Jet charge techniques have been 
used succesfully at LEP experiments for example for B-mixing 
measurements~\cite{jetcharge}.

In ATLAS, B$^{**}$ and jet charge tagging can be used to tag decays
${\rm B^0_d} \rightarrow$ J/$\psi$X, J/$\psi \rightarrow \mu^+\mu^-$, 
because the Level-1 trigger needs at least one muon passing the trigger
requirements. B$^{**}$ and jet charge tagging can thus be used in
order to increase statistics for the $\sin 2 \beta$ measurement. 
In addition, systematic effects related to tagging can be checked by
using the related decays ${\mathrm{B^0_d \rightarrow J/\psi K^{*0} }}$ and
${\mathrm{B^+ \rightarrow J/\psi K^+ }}$.

Only jet charge tagging can be used for tagging decays of the type
${\mathrm{B^0_s \rightarrow J/\psi X}}$, since 
there is no phase space for the decay $\mathrm{B^{**}_{u,d} \rightarrow
B^0_s K}$. Jet charge tagging can thus be used to 
increase statistics of $\gamma$ measurement with 
${\mathrm{B^0_s \rightarrow J/\psi \phi}}$, and possibly to
increase statistics of $x_{\mathrm s}$ measurement with
${\mathrm{B^0_s \rightarrow J/\psi K^{*0} }}$, if reconstruction of this
channel is otherwise feasible.
There are expected to be heavier $L$ = 2 states,
`$\mathrm{B^{***}}$' = $\mathrm{B^{*}_3(6148)} \rightarrow \mathrm{B_s K},
\mathrm{B_2(6148)} \rightarrow \mathrm{B^*_s K, B_s K^*}$~\cite{eichten}.
These states have not
been observed, but if these states were narrow and prominent, 
they could be useful for ${\mathrm{B^0_s}}$ tagging.

\boldmath
\subsection{B$^{**}$ tagging of ${\rm \bar{B}^0_d} \rightarrow$ J/$\psi$X, 
J/$\psi \rightarrow \mu^+\mu^-$}
\unboldmath

\begin{figure}[htb]
\begin{center}
\mbox{\epsfig{file=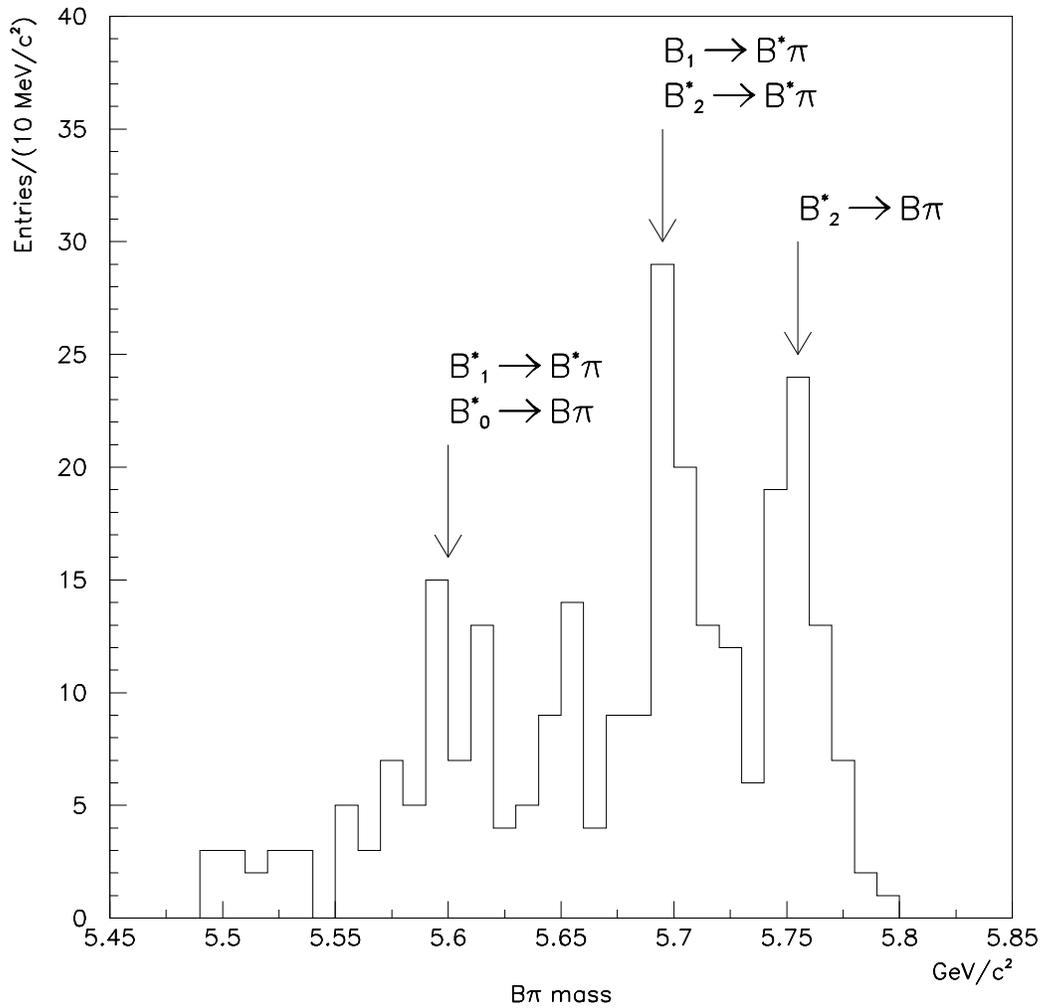,width=15.0cm}}
\caption[]{The B$\pi$ mass spectrum of the simulated B$^{**}$
sample. In B$^*$ final
states, the unreconstructed photon produces a mass shift of about 
46~MeV/$c^2$.}
\label{fig1}
\end{center}
\end{figure}

B$^{**}$ decays were implemented in PYTHIA 5.7, using a tuning based on ALEPH 
data, with production probability 
P(b$\rightarrow {\mathrm B}^{**}_{\mathrm{u,d}}$/b$\rightarrow
{\mathrm{B_{u,d}}}$)=0.3. The parton distribution function 
was the default one in PYTHIA 5.7, CTEQ2L.
The simulation study was performed at particle level 
without resolution smearings~\cite{patrik}. 
The B$\pi$ mass distribution of the simulated
B$^{**}$ sample is shown in Fig.~\ref{fig1}.

The selection of the non-tagged 
${\mathrm{\bar{B}^0_d \rightarrow J/\psi K^0_S }}$ 
sample was the same as for lepton analysis, apart from the tag requirements.
Normalizing to an integrated luminosity of $10^4$~pb$^{-1}$,
a sample of 59,000 reconstructed, 
untagged ${\mathrm{\bar{B}^0_d \rightarrow J/\psi K^0_S }}$ events was obtained
after the selection cuts.

The tag-pion was first required to be consistent with coming from the
primary vertex by requiring that the impact parameter was smaller
than 250~$\mu$m.
The tagging quality factor can be defined as
$Q_{\mathrm{tag}} = \varepsilon_{\mathrm{tag}} D_{\mathrm{tag}}^2$,
where $\varepsilon_{\mathrm{tag}}$ is the tagging efficiency. 
$Q_{\mathrm{tag}}$
was maximized by optimizing the minimum $p_{\rm T}$ of the pion,
maximum B$\pi$ invariant mass, maximum distance of the pion from the B,
$\Delta r = \sqrt{\Delta \eta^2 + \Delta \phi^2}$, and the maximum 
number of allowed pion candidates, $n_{\mathrm{cand}}$. 
To select the tag pion in case of multiple candidates passing the cuts,
a few selection algorithms were tried: selection based on 
1) maximum $p_{\rm L}$ with respect to the B, 
2) maximum $\cos \theta^*$, 
where $\theta^*$ is the angle between the pion
and the B$^{**}$ boost direction, after a boost to B$\pi$ rest frame, and
3) M(B$\pi$) closest to M(B$^{**}$).

\begin{figure}[htb]
\begin{center}
\mbox{\epsfig{file=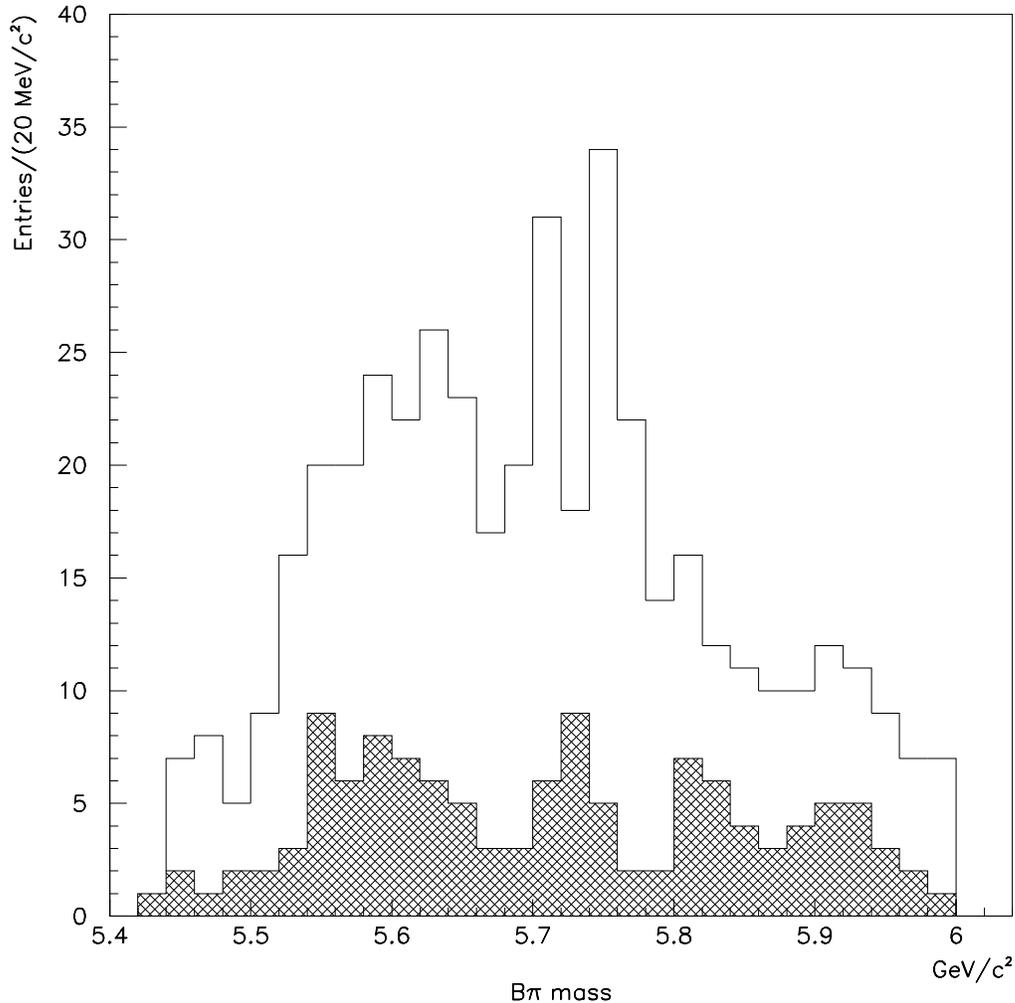,width=15.0cm}}
\caption[]{The B$\pi$ mass distribution of the tagged events.
The hatched histogram shows the wrong-sign combinations and the white
histogram shows the right-sign combinations. Of the 1688 events passing the
reconstruction cuts, 320 were tagged right and 122 wrong.
}
\label{fig2}
\end{center}
\end{figure}

Optimal cuts for the tag pion were found to be: 
$p_{\rm T} >$~1~GeV/$c$, $\Delta r < $~0.5, \\
M(B$\pi)<$~6.0~GeV/$c^2$,
$n_{\mathrm{cand}}$~=~1. Since events with only one tag-pion candidate 
were accepted, there was no need for additional selection criteria. 
The B$\pi$ mass distribution of the tagged events, using the optimized cuts,
is shown in Fig.~\ref{fig2}.
Results with optimal cuts were:
\begin{center}
\[ W_{\mathrm{tag}} = 0.28 \pm 0.02 \rightarrow D_{\mathrm{tag}} = 0.44 \]
\[ \varepsilon_{\mathrm{tag}} = 0.26 \pm 0.01 \]
\[ Q_{\mathrm{tag}} = 0.052 \pm 0.010 \]
\end{center}

Assuming the same S/B as for the lepton-tagged
sample, $\delta (\sin 2 \beta) (\mathrm{stat.}) = 0.032 \pm 0.003$.
When considering the channel 
J/$\psi \rightarrow \mu^+\mu^-$, the lepton-tag analysis showed that
the dominant background
originates from real J/$\psi$'s, since fake background is reduced by the 
high $p_{\rm T}$ cuts of the muons, and the good mass resolution. 
The lifetime cut of the B should suppress background from prompt J/$\psi$'s,
and work is in progress to verify this.

For comparison, the wrong-tag rate with lepton tags is 
$W_{\mathrm{tag}} = 0.22$, giving $D_{\mathrm{tag}} = 0.56$.
The lepton tagging efficiency includes trigger efficiencies, since the 
tag lepton is part of the trigger.
The theoretical maximum for the tagging efficiency would be about 0.20 (from
semileptonic branching ratios).

\boldmath
\subsection{ Jet charge tagging of 
        ${\mathrm{\bar{B}^0_d \rightarrow J/\psi K^0_S, J/\psi \rightarrow
         \mu^+\mu^-  }}$ }
\unboldmath

The simulation sample for the study of jet charge tagging was the same
as for the B$^{**}$ analysis, and
the same selection cuts were used~\cite{lena}.
Two different tagging methods were tried.
In same-side jet tagging, the tag was defined as the jet around the 
${\mathrm{J/\psi K^0_S}}$. The tagging variable was the
jet charge $Q_{\mathrm s}$.
A jet with $Q_{\mathrm s}<-c$ was defined as originating from a 
$\mathrm{\bar{B}^0_d}$, and if $Q_{\mathrm s}>c$, the jet was defined to 
originate from a $\mathrm{B^0_d}$.
In the same-side jet analysis, the jet cone size $\Delta r$, 
the $p_{\mathrm T}$ cut for the particles in the cone, and the
charge cut-value $c$ defining the flavour of the B 
were optimized to maximize the $Q_{\mathrm{tag}}$.

The other method studied was two-jet tagging, in which 
both the same side jet and the `other b-jet' were used for tagging.
The tagging variable was defined as the charge difference of the two jets, 
$dQ =  \langle Q_{\mathrm s} - Q_{\mathrm o} \rangle.$ 
Contrary to LEP, at LHC it is not obvious which is the other b-jet.
Here, the highest-$E_{\mathrm T}$ jet (excluding the same side jet)
was taken as the other b-jet
but the identification was correct only in 
25\% of the cases. We can, however, expect to improve this by optimizing the
jet reconstruction algorithm and including b-tagging for the other jet. 
For comparison, the other b-jet was also identified using 
the Monte Carlo simulation history in order to have an idea
of the optimal performance.

The jet charge was calculated as:
\begin{eqnarray}
  Q_{\mathrm{jet}} = \frac { \sum_{i=1}^n q_i (p_i)_{\mathrm T}^a } 
{\sum_{i=1}^n  (p_i)_{\mathrm T}^a }
\end{eqnarray}
where $(p_i)_{\mathrm T}$ is the transverse momentum of particle $i$ 
vs. beam direction.
Exponent $a$ was optimized along other parameters.
Algorithms based on longitudinal momentum $p_{\mathrm L}$ and 
rapidity $y$ along the jet axis were tried as well, 
with similar results.

\begin{figure}[htb]
\begin{center}
\mbox{\epsfig{file=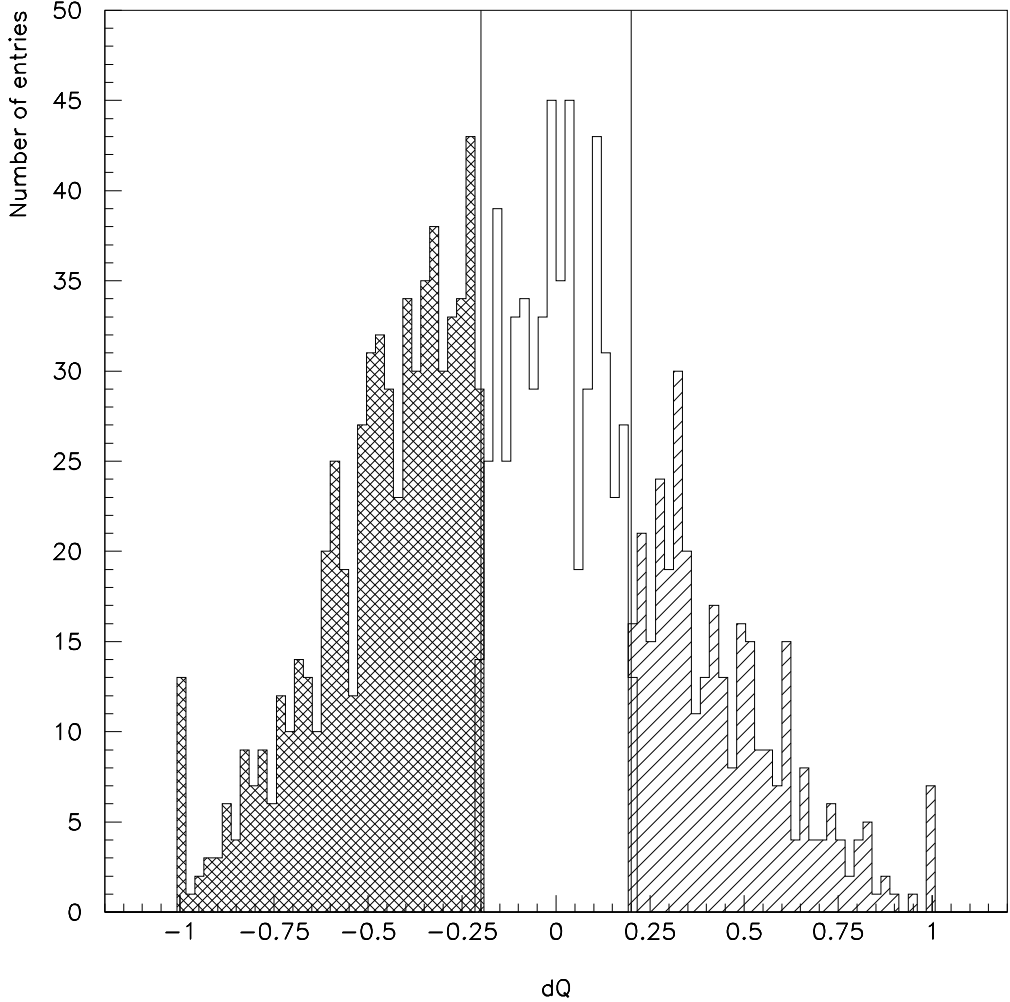,width=15.0cm}}
\caption[]{Distribution of the charge difference between the two jets, $dQ$.
Here, events with $dQ<-0.2$ are defined as tagged right (cross-hatched region) 
and jets with
$dQ>0.2$ are tagged wrong (hatched region), resulting in a wrong tag fraction
of about 34\%.
}
\label{fig3}
\end{center}
\end{figure}

Best results were obtained with
two-jet tagging using the $p_{\mathrm T}$-based algorithm for calculating
the jet charges.
The distribution of the tagging variable $dQ$ is shown in Fig.~\ref{fig3}.
The $Q_{\mathrm{tag}}$ was maximal with the following parameters:
$\Delta r$=0.7,  $p_{\mathrm T}>$~0.5~GeV/$c$, 
charge cut value $c$=0.4, exponent $a$=1.25.

The results were:
\begin{center}
\[ W_{\mathrm{tag}} = 0.34  \rightarrow D_{\mathrm{tag}} = 0.32 \]
\[ \varepsilon_{\mathrm{tag}} = 0.32 \]
\[ Q_{\mathrm{tag}} = 0.033 \pm 0.009 \]
\end{center}
Assuming the same S/B as for the lepton-tagged sample,
$\delta (\sin 2 \beta) (\mathrm{stat.}) = 0.040 \pm 0.003 $.

Similar results were obtained with
same-side jet tagging, using the rapidity-based algorithm for the jet-charge
calculation. With optimal parameters $\Delta r$=0.7, 
$p_{\mathrm T}>$~0.5~GeV/$c$, $c$=0.2, the tagging quality factor
$Q_{\mathrm{tag}}$ was 0.031.

For comparison, the result with ideal two-jet tagging (the other 
b-jet always found correctly) was:
$Q_{\mathrm{tag}} = 0.042 \pm 0.009$.

\subsection{Combined results}

Estimates of the ATLAS measurement accuracy of $\sin 2 \beta$ are summarized
in Tables~\ref{table3} and \ref{table4}. With the B$^{**}$-tagged sample, 
a statistical precision of $\pm$0.032 is obtained,
which is comparable to the precision obtained with the
muon-tagged sample with J/$\psi \rightarrow \mu^+\mu^-$ in the final state
($\pm$0.036).
There is a large overlap between
the event samples selected by the B$^{**}$ and jet charge tags, while the
overlap with the lepton-tagged sample and the other samples should be small.
Work is in progress to combine the different samples in an optimal way.

\begin{table}[htb]
\centering
\begin{tabular}{|l|c|c|}\hline
Parameter & $\mu$-tags & e-tags\\ \hline
$N_{\rm S}({\rm e^+e^- K^0_S})$ rec. & 9,540 & \\
$N_{\rm S}({\rm \mu^+\mu^- K^0_S})$ rec. & 6,220 & 5,360 \\ \hline
$N_{\rm S}$ & \multicolumn{2}{|c|}{21,120} \\ \hline
$N_{\rm B}({\rm {e^+e^-     \oplus \pi^+\pi^-}})$ & 3,820 & \\
$N_{\rm B}({\rm {\mu^+\mu^- \oplus \pi^+\pi^-}})$ & 120 & 200 \\ \hline
$\sqrt{D_{\rm {back}} }$ & \multicolumn{2}{|c|}{0.91} \\
$D_{\rm {tag}}$ & \multicolumn{2}{|c|}{0.56} \\
$D_{\rm {int}}$ & \multicolumn{2}{|c|}{0.63} \\ \hline
$\delta (\sin 2 \beta)$ (stat.) & \multicolumn{2}{|c|}{0.018 (time-dep.)}  \\ 
                                                        \hline
$\delta (\sin 2 \beta)$ (stat.) & \multicolumn{2}{|c|}{0.021 (time-int.)} 
                                      \\ 
$\delta (\sin 2 \beta)$ (syst.) & \multicolumn{2}{|c|}
  {$0.012 \cdot \sin 2 \beta \oplus 0.011$ (time-int.)}     \\ \hline
\end{tabular}
\caption{Summary on $\sin 2 \beta$ with 10$^4$ pb$^{-1}$: lepton tags.}
\label{table3}
\end{table}

\begin{table}[htb]
\centering
\begin{tabular}{|l|c|c|}\hline
Parameter & B$^{**}$-tags & Jet charge tags\\ \hline
$N_{\rm S}({\rm \mu^+\mu^- K^0_S})$ rec., untagged & 
                           \multicolumn{2}{|c|}{59,000} \\ \hline
$N_{\rm S}({\rm \mu^+\mu^- K^0_S})$ rec., tagged & 15,340 & 18,760 \\ \hline
$\sqrt{D_{\rm {back}} }$ (assumed) & \multicolumn{2}{|c|}{0.91} \\ \hline
$D_{\rm {tag}}$ & 0.44 & 0.32 \\ \hline
$D_{\rm {int}}$ & \multicolumn{2}{|c|}{0.63} \\ \hline
$\delta (\sin 2 \beta)$ (stat.) & 0.032 (time-int.) & 0.040 (time-int.)
                                                               \\ \hline
\end{tabular}
\caption{Summary on $\sin 2 \beta$ with 10$^4$ pb$^{-1}$: B$^{**}$ and 
jet charge tags.}
\label{table4}
\end{table}

\section{Other CP-violation measurements}

ATLAS capabilities of measuring $\sin 2\alpha$ have been investigated
in~\cite{tp,pipi}. With an integrated luminosity of 10$^4$ pb$^{-1}$, 
7,120 signal events are expected, with a background of 6,475 events from
other exclusive two- or three-body B-decays, and 565 events from combinatorial
background. Neglecting the penguin contributions to the decay, the statistical
error on $\sin 2\alpha$ has been estimated to be $\pm0.043$.
The experimental systematic uncertainty is dominated by the large background,
since ATLAS does not have charged hadron identification.
Assuming a 20\% systematic uncertainty for the level of background,
the systematic uncertainty of $\sin 2\alpha$ is 
$\pm 0.10 \cdot \sin 2 \alpha \oplus 0.011$.

Since ATLAS is not expected to be able to reconstruct decays 
${\mathrm{B^0_d \rightarrow \pi^0 \pi^0}} $ and \\
${\mathrm{B^\pm \rightarrow \pi^\pm \pi^0}} $,
ATLAS cannot measure the parameter
$\rho = A_P/A_T$, which is the ratio of penguin to tree amplitudes.
If $\rho$ is measured elsewhere, or calculated reliably, 
ATLAS should be able to measure $\sin 2 \alpha$ despite of the penguin effects.
If one pessimistically assumes $\rho = 0.2 \pm 0.2$, the theoretical 
systematic uncertainty can be very large, 
$\delta \sin 2 \alpha \simeq$~0.1-0.3. 

In the decay channel ${\rm B^0_s \rightarrow J/\psi \phi}$, a small asymmetry
is expected in the Standard Model,
\begin{eqnarray}
A(t) \propto D \cdot 2 |V_{cd}||V_{ub}/V_{cb}| \cdot 
 \sin \gamma \sin \Delta m_{\rm s} t \simeq 0.03 \cdot D \cdot 
 \sin \gamma \sin \Delta m_{\rm s} t .
\end{eqnarray}
New physics phenomena, contributing to B-mixing, could enhance
the asymmetry substantially.

A study of this decay channel was documented in~\cite{tp,bc}.
With an integrated luminosity of 10$^4$ pb$^{-1}$,
a signal of 15,000 lepton-tagged events is expected, with a
background of 3,000 events. The precision with which the asymmetry
can be measured depends on the value of $x_{\mathrm s}$,
because a time-dependent measurement of the asymmetry is mandatory due
to the expected large value of $x_{\mathrm s}$. Assuming $x_{\mathrm s}$ = 10,
$\delta(A_{\rm obs}) = 0.025$, and with $x_{\mathrm s}$ = 25,
$\delta(A_{\rm obs}) = 0.06$. 
The signal statistics could be increased with jet charge tagging.

\boldmath
\section{Measurement of $x_{\mathrm s}$}
\unboldmath

In addition to the angles of the unitarity triangle, measurements of the
lengths of the sides provide us with complementary information about the
triangle. The length of the side $|V_{td}|$ is least well-known of the sides.
The mixing parameter $x_{\rm d}$ is proportional to $|V_{td}|^2$,
but inferring the value of $|V_{td}|^2$ from $x_{\rm d}$
is hampered by hadronic uncertainties.
To large extent, these uncertainties
cancel, when considering the ratio of the mixing parameters, 
$x_{\rm s}/x_{\rm d} \propto |V_{ts}/V_{td}|^2$.

The mixing parameter $x_{\rm s}$ has not been measured yet. The Standard Model
predicts it to be in the range 10-30, and the present lower limit from LEP
is $x_{\rm s} > 11.1$~\cite{zeitnitz}.

The following decay channels were studied for $x_{\rm s}$ measurement
in ATLAS:
$\mathrm{ B_s^0 \rightarrow  D_s^- \pi^+}$, 
$\mathrm{D_s^- \rightarrow \phi^0  \pi^-}$,
$\mathrm{\phi^0 \rightarrow K^+ K^-} $
and
$\mathrm{ B_s^0 \rightarrow  D_s^- a_1^+}$, 
$\mathrm{D_s^- \rightarrow \phi^0  \pi^-}$,
$\mathrm{\phi^0 \rightarrow K^+ K^-}$, 
$\mathrm{a_1^+ \rightarrow \rho^0 \pi^+}$,\\
$\mathrm{\rho^0 \rightarrow \pi^+\pi^-}$.

\boldmath
\subsection{$\mathrm{ B_s^0 \rightarrow  D_s^- \pi^+}$}
\unboldmath

The first level trigger for this decay channel is a single muon
coming from the semileptonic decay of the accompanying b.
The trigger muon also serves as a tag
defining the flavour of the $\mathrm{ B_s^0}$ at production. The
second level trigger is a $\mathrm{  D_s}$ mass trigger.

Three-dimensional vertex fits of 
$\mathrm {D_s^-}$ and $\mathrm {B_s^0}$ were performed~\cite{bs}. 
From full simulation, the decay time resolution of the B was found to be
$\sigma_t({\rm B})=0.069 \pm 0.008 \ \ \mathrm{ps}$. The
$\mathrm {B_s^0}$ mass resolution was 33 MeV/$c^2$, and the 
$\mathrm {D_s^-}$ mass resolution 10 MeV/$c^2$.

After the reconstruction cuts, and including trigger and reconstruction 
efficiencies, 3,640 reconstructed events were obtained, assuming an 
integrated luminosity of $10^4$~pb$^{-1}$.

Backgrounds studied were:
related B decays ${\mathrm {B^0_d \rightarrow D^-_s} \pi^+}$, 
${\mathrm {B^0_d \rightarrow D^- \pi^+}}$, and
${\mathrm {\Lambda_b \rightarrow \Lambda^+_c \pi^-}}$,
${\mathrm {\Lambda^+_c \rightarrow p K^- \pi^+ }}$.
Combinatorial background was checked as well.
No significant background from these sources was found, and in the 
consequent analysis, the signal-to-background ratio was assumed to be 
1.6, which should be an upper limit for the background.

\boldmath
\subsection{$\mathrm{ B_s^0 \rightarrow  D_s^- a^+_1(1260)}$}
\unboldmath

The trigger for this channel is as for the decay
$\mathrm{ B_s^0 \rightarrow  D_s^- \pi^+}$.
Three-dimensional 
vertex fits were performed to reconstruct 
$\mathrm{D_s^-}$, $\mathrm{a_1^+}$ and $\mathrm{B_s^0}$~\cite{bsa1}.
To find $\mathrm{a_1^+}$, further cuts were applied on
$\Delta \phi_{\pi \pi}$~,~ $\Delta \Theta_{\pi \pi}$~,~
$|M_{\pi \pi}-M_\rho|$  and $|M_{\pi \pi \pi}-M_{\mathrm{a_1}}|$. 
The decay time resolution was found to be 0.064 ps.

No significant background was found from:
${\mathrm {B^0_d \rightarrow D^-_s} a^+_1}$,
${\mathrm {B^0_d \rightarrow D^- a^+_1}}$,
${\mathrm {\Lambda_b \rightarrow \Lambda^+_c \pi^-}}$,
${\mathrm {\Lambda^+_c \rightarrow p K^- \pi^+ \pi^-\pi^+ }}$, and
combinatorial background.

After the reconstruction cuts, and including trigger and reconstruction 
efficiencies, 1,230 reconstructed events were obtained.

\boldmath
\subsection{The $x_{\rm s}$ reach}
\unboldmath

The asymmetry between the decay time distributions of 
unmixed $(++)$ and mixed B's $(+-)$ is:
\begin{eqnarray}
A(\tau) = \frac{dn(+ +)/d\tau - dn(+ -)/d\tau}
             {dn(+ +)/d\tau + dn(+ -)/d\tau}
       = D \cos (x_{\mathrm s} \tau/\tau_{\mathrm{Bs}}) , 
\end{eqnarray}
where $D$ is the product of all dilution factors,
$D_{\mathrm{tag}} = 0.56$,
$D_{\mathrm{back}} = 0.61$ (S/B=1.6), $D_{\mathrm{time}}$ (dependent on
$x_{\mathrm s}$).

In~\cite{tp}, we used the power spectrum of Fourier transform to obtain the 
value of $x_{\mathrm s}$ from the asymmetry. This kind of transformation can be
applied to any type of asymmetry function.

A refined fitting method, an amplitude fit, has been proposed in~\cite{moser}.
The amplitude fit is equivalent to 
using a Fourier cosine transform. In this method, 
a cosine-function, or an even function in more general case, 
is explicitly required.

The amplitude fit is performed in the following way:
$x_{\rm s}$ values are scanned, and for every fixed $x_{\rm s}$,
the amplitude $D$ in the expression
$D \cos (x_{\rm s} \tau/\tau_{\mathrm{Bs}})$ is fitted. 
The ${\mathrm{B^0_s}}$ lifetime is fixed to its measured value.
The $x_{\rm s}$ value which gives the 
maximum fitted amplitude $D$, is then the measured value.

\begin{figure}[htb]
\begin{center}
\mbox{\epsfig{file=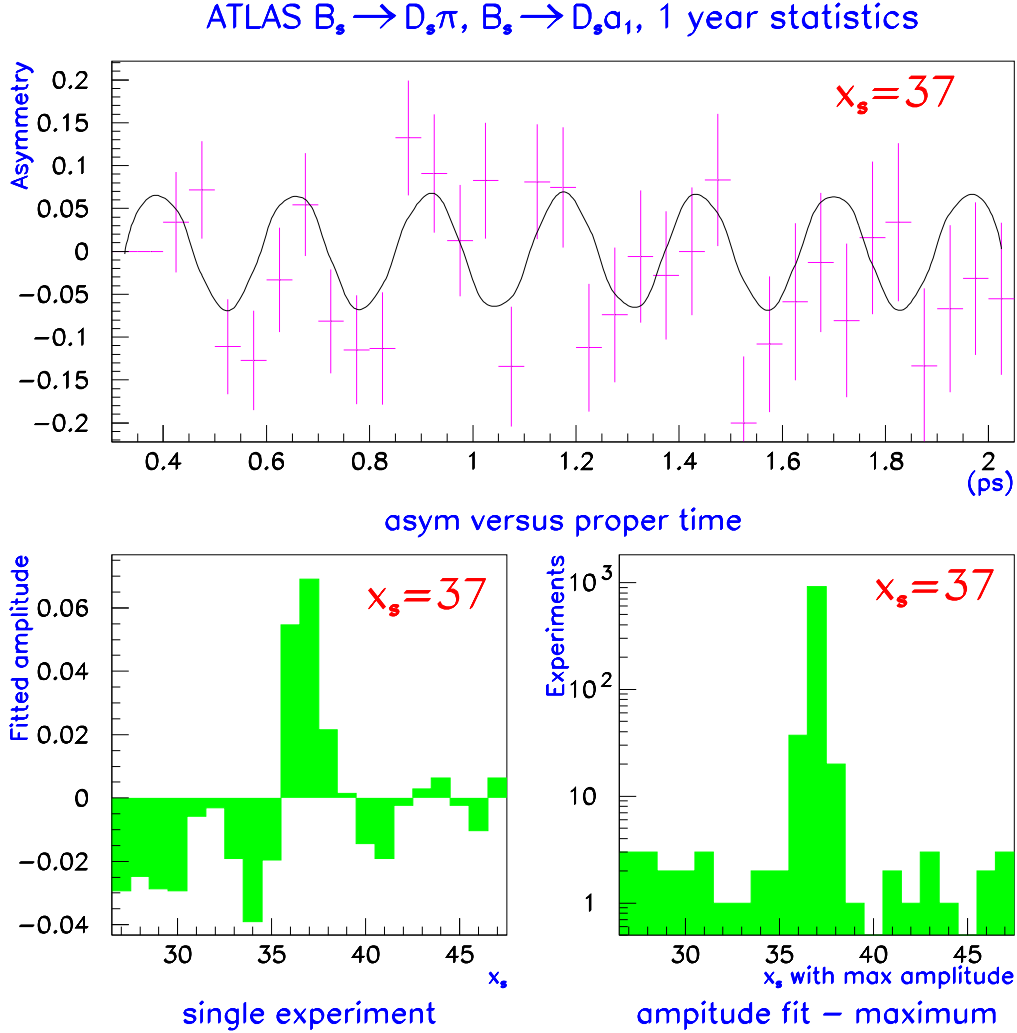,width=15.0cm}}
\caption[]{a) Asymmetry versus proper time for $x_{\mathrm s}$=37.
b) Result of the amplitude fit with  $x_{\mathrm s}$ scan.
c) Distribution of the $x_{\mathrm s}$ corresponding to the maximum amplitude,
when the experiment was repeated 1,000 times.
}
\label{fig4}
\end{center}
\end{figure}

The reachable $x_{\rm s}$ range was defined by repeating the
`experiment' 1,000 times, and in 95\% of the cases,
a correct answer was required from the amplitude fit.
Combining the two decay modes, it was found that ATLAS can measure 
$x_{\rm s}$ values up to 37 with one year's statistics (10$^4$~pb$^{-1}$).
The results are shown in Fig.~\ref{fig4} for samples generated with
$x_{\rm s}=37$.

Finally, ATLAS will also be able to measure $x_{\rm d}$ with a good 
precision, and thus we can measure independently the ratio 
$x_{\rm s}/x_{\rm d}$.

\boldmath
\section{Rare decays $\mathrm{B \rightarrow \mu^+\mu^-(X)}$}
\unboldmath

LHC provides a sufficient $\mathrm{b\bar{b}}$ production rate to 
study rare decays $\mathrm{B \rightarrow \mu^+\mu^-(X)}$. The decays 
are easily triggered with a low-rate di-muon trigger. 
The analysis cuts are based on
secondary vertex requirements and B decay kinematics~\cite{flora1,flora2}. 

With an integrated luminosity of $10^4$~pb$^{-1}$,
the upper limit to the branching ratio of
the decay ${\rm B^0_{s} \rightarrow \mu^+ \mu^-}$ can be established 
to be BR(${\rm B^0_s \rightarrow \mu^+ \mu^-}$) = 
2.5~$\times$ 10$^{-9}$ at 95\%~C.L..
The Standard Model predicts a branching ratio ${\cal O}(10^{-9})$.
After two low-luminosity LHC-years, a two-standard-deviation
signal can be seen.

For the decay ${\rm B^0_{d} \rightarrow \mu^+ \mu^-}$, the 
Standard Model prediction for the branching ratio is ${\cal O}(10^{-10})$.
An upper limit of
BR(${\rm B^0_d \rightarrow \mu^+ \mu^-}$) = 
6.6~$\times$ 10$^{-10}$ can be established at 95\%~C.L..

The ratio of the branching ratios of the decays
$\rm{B^0_d} \rightarrow \rho^0 \mu^+ \mu^-$ and
$\rm{B^0_d} \rightarrow \rm{K^{*0}} \mu^+ \mu^-$ provides us with
an independent measurement of the CKM matrix elements:
\begin{eqnarray}
\frac{\rm{Br}(\rm{B^0_d} \rightarrow \rho^0 \mu^+ \mu^-)}
  {\rm{Br}(\rm{B^0_d} \rightarrow \rm{K^{*0}} \mu^+ \mu^-)}  =
 k_d \frac{|V_{td}|^2}{|V_{ts}|^2},
\end{eqnarray}
where $k_d = (F^{\rm{B^0_d}\rightarrow \rho^0}/F^{\rm{B^0_d}\rightarrow
 \rm{K^{*0}}})^2$ is the ratio of form factors squared.
Ratio of the form factors 
is calculable with an error of 5 - 10\%. 

In the decay channel
${\rm B^0_{d} \rightarrow K^{*0} \mu^+ \mu^-}$, a signal
of 2,450 events is expected assuming the Standard Model 
branching ratio, over a combinatorial background of 420 events.
In the decay channel ${\rm B^0_{d} \rightarrow \rho^0 \mu^+ \mu^-}$,
330 signal events (produced according to the branching ratio predicted by
the Standard Model) are expected to be observed
over a background of 860 events, consisting of both combinatorial background
and reflection from 
${\rm B^0_{d} \rightarrow K^{*0} \mu^+ \mu^-}$.

\section{Other B-physics topics with ATLAS}

Among other B-physics topics, B-baryons and ${\rm{B_c}}$ mesons have been
studied in ATLAS. The polarization of
$\Lambda_{\rm b}$ is expected to be measured with a statistical 
accuracy of 0.008, using the decay mode J/$\psi \Lambda$~\cite{maria}.
The $\Xi_{\rm b}$ polarization will be measured with a statistical 
accuracy of 0.14, using the same decay mode.
In decay mode ${\rm{B_c}} \rightarrow$~J/$\psi \pi$, 
a signal of about 4,000 events
with a 14-standard-deviation significance can be established 
with $10^{4}$~pb$^{-1}$ of data~\cite{albiol}.

\section{Conclusions}

ATLAS will be very competitive in the measurement of
$\sin 2 \beta$, due to the possibility to use low-$p_{\rm T}$ electrons,
good vertex resolution, and versatile tagging methods.
In addition, ATLAS will contribute to the measurements of other angles
of the unitarity triangle: angles $\alpha$ and $\gamma$. 

ATLAS will be able to measure a side of the triangle 
independently. In addition to the measurement of $x_{\rm d}$, the
measurement of $x_{\rm s}$ will be possible by utilizing
the dedicated vertex layer close to the beam pipe.

ATLAS will have access to
rare decays ${\mathrm{B \rightarrow \mu^+\mu^- (X)}}$, due to the efficient
muon trigger and identification, and vertexing.
These decays provide us with a Standard
Model (in)consistency test.
ATLAS will also have access to rare B-hadrons such as ${\mathrm{B_c}}$ and
baryons with more than one heavy quark, which will test
heavy quark models.

\section{Acknowledgement}
Special thanks to CERN 1995 summer students Patrik Johnsson and Lena Leinonen
for their contributions to the results presented in this paper.

\end{document}